\documentclass[12pt,a4paper,english,english,english]{article}
\usepackage[T1]{fontenc}
\usepackage[latin1]{inputenc}
\usepackage{babel}

\makeatletter

\providecommand{\LyX}{L\kern-.1667em\lower.25em\hbox{Y}\kern-.125emX\@}

\usepackage[T1]{fontenc}
\usepackage[latin1]{inputenc}
\usepackage{babel}

\makeatletter

\usepackage[T1]{fontenc}
\usepackage[latin1]{inputenc}
\usepackage{babel}
\usepackage{setspace}
\makeatletter

\usepackage[T1]{fontenc}
\makeatletter

\usepackage[T1]{fontenc}

\makeatother

\makeatother

\makeatother

\makeatother
\begin{document}
\baselineskip .3in

\begin{titlepage}

{\noindent \centering \textbf{\large Precursors of catastrophe in
the Bak-Tang-Wiesenfeld, Manna and random fiber bundle models of failure }\large \par}

{\noindent \centering \vskip .3in\par}

{\noindent \centering \textbf{Srutarshi Pradhan} \( ^{(1)} \)and
\textbf{Bikas K. Chakrabarti} \( ^{(2)} \)\par}

{\noindent \centering \textit{Saha Institute of Nuclear Physics},\\
 \textit{1/AF Bidhan Nagar, Kolkata700 064, India.}\\
\par}

\noindent \vskip .3in

\noindent \textbf{Abstract}

\noindent We have studied precursors of the global failure in some
self-organised critical models of sand-pile (in BTW and Manna models)
and in the random fiber bundle model (RFB). In both BTW and Manna
model, as one adds a small but fixed number of sand grains (heights)
to any central site of the stable pile, the local dynamics starts
and continues for an average relaxation time \( \tau  \) and an average
number of topplings \( \Delta  \) spread over a radial distance \( \xi  \).
We find that these quantities all depend on the average height \( h_{av} \)
of the pile and they all diverge as \( h_{av} \) approaches the critical
height \( h_{c} \) from below: \( \Delta  \) \( \sim (h_{c}-h_{av}) \)\( ^{-\delta } \),
\( \tau \sim (h_{c}-h_{av})^{-\gamma } \) and \( \xi  \) \( \sim  \)
\( (h_{c}-h_{av})^{-\nu } \). Numerically we find \( \delta \simeq 2.0 \),
\( \gamma \simeq 1.2 \) and \( \nu \simeq 1.0 \) for both BTW and
Manna model in two dimensions. In the strained RFB model we find that
the breakdown susceptibility \( \chi  \) (giving the differential
increment of the number of broken fibers due to increase in external
load) and the relaxation time \( \tau  \), both diverge as the applied
load or stress \( \sigma  \) approaches the network failure threshold
\( \sigma _{c} \) from below: \( \chi  \) \( \sim  \) \( (\sigma _{c} \)
\( - \)\( \sigma )^{-1/2} \) and \( \tau  \) \( \sim  \) \( (\sigma _{c} \)
\( - \)\( \sigma )^{-1/2} \). These self-organised dynamical models
of failure therefore show some definite precursors with robust power
laws long before the failure point. Such well-characterised precursors
should help predicting the global failure point of the systems in
advance.\\
 \vskip .3in

\noindent \textit{e-mail addresses} :

\noindent \( ^{(1)} \)spradhan@cmp.saha.ernet.in

\noindent \( ^{(2)} \)bikas@cmp.saha.ernet.in

\noindent \end{titlepage}

\noindent \textbf{I. Introduction }

\noindent In a sand-pile, whenever the local slope at the surface
of the pile exceeds the angle of repose, avalanches take place and
the sand grains move to the neighbouring sites. If the local slope
of these neighbouring sites increase, in turn, beyond the angle of
repose, avalanches continue. Otherwise the dynamics stops until another
sand grain is added to the pile. The system finally attains a self-organised
state where extra grains, when added, get out of the system through
successive avalanches from its boundary. Models of sand-piles have
been developed to study such self-organisation. Bak, Tang and Wiesenfeld
(BTW) \cite{BTW187,BTW287} introduced the random height sand-pile
model where height units are added randomly at any site at a constant
rate and a site topples when its height equals an integer threshold
value \( h_{th} \) (= 4 for square lattice, for example). Whenever
any site topples, the local height becomes zero there and the height
is locally conserved by equal sharing among the nearest neighbours
(four in number for square lattice) and the neighbours get one unit
of height added to theirs'. The boundary of the system is completely
absorbing. As more and more grains (heights) are added slowly to the
system, the average height \( h_{av} \) of the system gradually increases
and attains a critical height \( h_{c} \) (equivalent to the angle
of repose of the sand-pile), beyond which the growth of average height
stops as the further addition of grains at any site causes successive
avalanches or failures of all sizes. These happen due to the long-range
correlations developed and the additional grains finally get out of
the system through its boundaries. The self-organised state here becomes
critical as it involves power law behaviour in avalanche size distribution
and the corresponding lifetime distribution. Extensive numerical checkings
confirmed this self-organised critical behaviour in both two and three
dimensional sand piles \cite{GM90,B2000}. Later, Manna introduced
\cite{Man91} a two state and stochastic version of the BTW model
where the threshold height has been chosen to be two (\( h_{th}=2 \)).
The toppling at any site reduces the height there to zero and the
toppled heights add to the height of any stocastically chosen site
among the four neighbouring sites of the toppled one. Here also, with
constant addition of sand grains, the system gradually reaches again
a critical state and there the avalanche size distribution and the
corresponding lifetime distribution again follow similar scaling behaviour.
However, the exponents for the Manna model seem to be different \cite{Man91,DD99}
from those of the BTW model. A similar self-organising dynamics is
also seen in a strained random fiber bundle (RFB) model \cite{Dan45,HH92, HH94,D92,RS99,Pach2000}
where \( N \) fibers are connected in parallel to each other and
clamped between their two ends. The strength of the individual fibers
has a random distribution (white, Gaussian or otherwise). Under a
load \( F \), a fraction of the fibers fail immediately whose strengths
are less than the stress \( \sigma  \) (\( =F/N \)). After this,
the total load of the bundle redistributes globally as the stress
is transferred from broken fibers to the remaining unbroken ones.
This redistribution causes secondary failures which in general causes
further failures and so on. After some typical relaxation time \( \tau  \)
(dependent on \( \sigma  \)), the system ultimately becomes stable
if the applied stress \( \sigma  \) is less than a critical value
\( \sigma _{c} \), beyond which all the fibers break and the network
fails completely. Although the RFB model is not a self-organised critical
one (as the failure state at \( \sigma  \) \( > \) \( \sigma _{c} \)
is not critical), it has some self-organising dynamics (stress redistribution
for \( \sigma  \) \( \leq  \) \( \sigma _{c} \)) similar to the
earlier ones and is very simple to tackle analytically. The studies
of these self-organising model systems and their scaling behaviour
have been extremely useful in analysing the statistics of fracture
and breakdown in real materials, including in earthquakes \cite{B2000,BB97,DT99}.

An obvious question arises: Are there any precursors or prior indications
which can tell how far a (slowly) growing sand-pile or a gradually
strained fiber bundle is away from its global failure point? The study
of precursors in self-organised systems was initiated by Acharyya
and Chakrabarti \cite{AC96}. Here, the global failure is identified
as the system spanning avalanche occurring at \( h_{av}=h_{c} \).
They tried to study the response of BTW model to pulsed addition of
grains (heights) in two and three dimensional sand-piles, where, `pulse'
means a fixed number of grains, added at any site to trigger the dynamics
locally in time and space. Adding a pulse of heights at any site of
a stable pile (where toppling had stopped), they measured the response
of the system in terms of the number of affected or toppled sites
(\( \Delta  \)) and the corresponding response or relaxation time
(\( \tau  \)) at various average heights (\( h_{av} \)) of the system.
They observed that both \( \Delta  \) and \( \tau  \) diverge as
\( h_{av} \) approaches the critical height \( h_{c} \). They also
estimated the exponents involved in the power laws for these divergences.
However, these estimates for the exponent values were not quite accurate
due to the small system sizes considered and strong pulses applied.
Similarly, the breakdown susceptibility \cite{AC96} of the RFB model
was studied by measuring the increment in the number of broken fibers
with the increment in the stress \( \sigma  \) \cite{PS97}. It was
seen that this differential increase in the number of broken fibers
due to infinitesimal increase in stress \( \sigma  \), diverges as
the stress \( \sigma  \) approaches the global failure threshold
\( \sigma _{c} \).

In this paper, we have studied several precursors in the models of
sand-piles and random fiber bundle. We have studied the response of
sand-pile models (both BTW and Manna model) to pulsed addition of
sand grains (heights; for unit time or unit pulse width), where the
applied pulse strength is negligible, so that the statistical state
of the system is not perturbed significantly by the applied pulse.
We have identified three parameters, namely, the total number of topplings
(\( \Delta  \)), the corresponding relaxation time (\( \tau  \))
and the correlation length (\( \xi  \)); all of which diverge as
the average height (\( h_{av} \)) of the pile approaches the critical
height (\( h_{c} \)). The values of the exponents for the variations
of these quantities (\( \Delta  \), \( \tau  \) and \( \xi  \))
with \( h_{av} \) near \( h_{c} \) have been estimated accurately.
In fact, the estimated value of the critical height or the location
of the catastrophe point \( h_{c} \), extrapolated separately from
the growing (precursor) values of \( \Delta  \), \( \tau  \) and
\( \xi  \) (for \( h_{av} \) values below \( h_{c} \)), agree quite
well with the previous direct numerical estimates \cite{Manna90}
for the same. In the RFB model, we have studied the breakdown susceptibility
(\( \chi  \)) and the response time (\( \tau  \)) required for the
bundle to become stable when an initial load or stress \( \sigma  \)
(\( <\sigma _{c} \)) is applied on it. Both \( \chi  \) and \( \tau  \)
diverge as \( \sigma  \) approaches \( \sigma _{c} \). The growth
behaviour of these precursors for \( \sigma  \) below \( \sigma _{c} \)
and the possibility of their extrapolations for estimating the failure
point \( \sigma _{c} \) of the network is discussed.

\vskip .2in

\noindent \textbf{II. Precursors in the BTW model}

\noindent \textbf{a) Model}

Let us consider a BTW model on a square lattice of size \( L\times L \).
At each lattice site \( (i,j) \), there is an integer variable \( h_{i,j} \)
which represents the height of the sand column at that site. A unit
of height (one sand grain) is added at a randomly chosen site at each
time step and the system evolves in discrete time. The dynamics starts
as soon as any site \( (i,j) \) has got a height equal to the threshold
value (\( h_{th} \)= \( 4 \)): the site topples, i.e., \( h_{i,j} \)
becomes zero there, and the heights of the four neighbouring sites
increase by one unit \begin{equation}
\label{00}
h_{i,j}\rightarrow h_{i,j}-4,h_{i\pm 1,j}\rightarrow h_{i\pm 1,j}+1,h_{i,j\pm 1}\rightarrow h_{i,j\pm 1}+1.
\end{equation}
 If, due to this toppling at site \( (i,j) \), any neighbouring site
become unstable (its height reaches the threshold value), they in
turn follow the same dynamics. The process continues till all sites
become stable (\( h_{i,j}< \) \( h_{th} \) for all \( (i,j) \)).
When toppling occurs at the boundary of the lattice (four nearest
neighbours are not available), extra heights get off the lattice and
are removed from the system.

With a very slow but steady rate of addition of unit height (sand
grain) at random sites of the lattice, the avalanches get correlated
over longer and longer ranges and the average height (\( h_{av} \))
of the system grows with time. Gradually the correlation length (\( \xi  \))
becomes of the order the system size \( L \). Here, on average, the
additional height units start leaving the system as the system approaches
toward a critical average height \( h_{c}(L) \) and the average height
remains stable there (see Fig. 1). Also the system becomes critical
here as the distributions of the avalanche sizes and the corresponding
life times follow robust power laws \cite{GM90,B2000}. In fact, a
finite size scaling fit \( h_{c}(L)=h_{c}(\infty )+{\textrm{C}}L^{-1/\nu } \)
(obtained by setting \( \xi  \) \( \sim  \) \( \mid h_{c}(L)-h_{c}(\infty )\mid ^{-\nu }=L \)),
where C is a constant, with \( \nu \simeq 1.0 \) gives \( h_{c}\equiv h_{c}(\infty )\simeq 2.124 \)
(see inset of Fig. 1). Similar finite size scaling fit with \( \nu =1.0 \)
gave \( h_{c}(\infty )\simeq 2.124 \) in earlier large scale simulations
\cite{Manna90}.

\noindent \textbf{b) Simulation studies for pulsed perturbation}

We have taken random height BTW systems on square lattice of different
sizes (\( L= \) \( 100 \), \( 200 \) and \( 300 \)). At a fixed
value of \( L \), for any pile configuration at an average height
\( h_{av} \), when all sites of the system have become stable (dynamics
have stopped), a fixed number of height units \( h_{p}=4 \) (pulse
of sand grains) is added at any central point of the system. Just
after this addition, the local dynamics starts and it takes a finite
time or iterations to return back to the stable state (\( h_{i,j}<4 \)
for all \( (i,j) \)) after several toppling events. For each value
of \( h_{av}(<h_{c}) \), we take about \( 10^{5} \) initial configurations
and this response or relaxation time has been noted for each of them.
The average relaxation time \( \tau  \) is obtained taking averages
over all configurations and is seen to diverge as \( h_{av} \) approaches
the critical height \( h_{c} \) (see Fig. 2 (a)). Near \( h_{c} \),
\( \tau  \) follows a power law \( \tau \sim (h_{c}-h_{av})^{-\gamma } \),
where \( \gamma \cong 1.2 \). The plot of \( \tau ^{-1/\gamma } \)
with \( h_{av} \) is a straight line with negative slope. Extrapolating
the straight line and locating the vanishing point of \( \tau ^{-1/\gamma } \)
one can estimate the critical point \( h_{c}=2.13\pm .01 \) (see
inset of Fig. 2 (a)) which is very close to the previous numerical
estimate \( h_{c}\cong 2.124 \) \cite{Manna90}.

Another response parameter, the average size of the damage (\( \Delta  \)),
i.e., the average number of topplings (after the addition of pulse)
has been measured as follows: the number of topplings for each configuration
at each value of \( h_{av} \) is noted and averaged out over the
initial configurations (about \( 10^{5} \) in number). Thus the average
\( \Delta  \) for that value of \( h_{av} \) is estimated and this
is also seen to diverges as \( h_{av}\rightarrow  \) \( h_{c} \)
(see Fig. 2 (b)). Near the critical point, we find \( \Delta  \)
\( \sim (h_{c}-h_{av}) \)\( ^{-\delta } \), where \( \delta \cong 2.0 \).
The plot of \( \Delta  \)\( ^{-1/\delta } \) versus \( h_{av} \)
gives a straight line with negative slope (see inset of Fig. 2 (b))
which can again be used to estimate \( h_{c} \) \( (=2.12\pm .01) \)
after extrapolating the straight line up to the vanishing point of
\( \Delta  \)\( ^{-1/\delta } \).

We have also measured the correlation length \( \xi  \) of the system
during the same experiment. When the pulse is added at any central
point \( (i_{0},j_{0}) \) of the system at some \( h_{av} \), toppling
starts there and gradually it moves toward the boundaries . We have
marked the farthest affected site \( (i_{f},j_{f}) \) (where at least
one toppling has occurred due to the pulse) with respect to the central
site \( (i_{0},j_{0}) \) where the pulse had been added. Clearly,
the average (over configurations) distance between the central and
the farthest affected sites \( (| \)\( (i_{0},j_{0})- \)\( (i_{f},j_{f})|) \)
is a measure of the correlation length of the system at that \( h_{av} \).
This correlation length \( \xi  \) is seen to diverge as \( h_{av}\rightarrow  \)
\( h_{c} \) (see Fig. 2 (c)) following a power law \( \xi  \) \( \sim  \)
\( (h_{c}-h_{av})^{-\nu } \), where \( \nu \cong 1.0 \). The plot
of \( \xi  \)\( ^{-1/\nu } \) versus \( h_{av} \) (see inset of
Fig. 2 (c)) is a straight line. The vanishing point of \( \xi  \)\( ^{-1/\nu } \)
gives an estimate of the critical point \( h_{c} \) and we find \( h_{c} \)
\( =2.13\pm .01 \). This is also close to the previously estimated
critical value.

\vskip .2in

\noindent \textbf{III. Precursors in the Manna model}

\noindent \textbf{a) Model}

We consider now the Abelian Manna model on a square lattice of size
\( L\times L \), where the sites can be either empty or occupied
with unit height i.e., the height variables can have binary states
\( h_{i,j}=1 \) or \( h_{i,j}=0 \). A site is chosen randomly and
one height is added at that site. If the site is initially empty,
it gets occupied: \begin{equation}
\label{pp}
h_{i,j}\rightarrow h_{i,j}+1,
\end{equation}
 If the chosen site is previously occupied then a toppling or `hard
core interaction' rejects both the heights from that site: \begin{equation}
\label{qq}
h_{i,j}\rightarrow h_{i,j}-2,
\end{equation}
 and each of these two rejected heights stochastically chooses its
host among the 4 neighbours of the toppled site. The toppling can
happen in chains if any chosen neighbour was previously occupied and
thus cascades are created. After the system attains stable state (dynamics
stopped), a new site is chosen randomly and unit height is added to
it. Thus the system evolves in discrete time steps. Here again the
boundary is assumed to be completely absorbing so that heights can
leave the system due to the toppling at the boundary.

With a slow rate of addition of heights at random sites, initially
the average height of the system grows with time and soon the system
approaches toward a critical average height \( h_{c} \), where the
average height stabilizes and does not change with further addition
of heights (see Fig. 3). The critical average height \( h_{c} \)
has a finite size dependence and a similar finite size scaling fit
\( h_{c}(L)=h_{c}(\infty )+{\textrm{C}}L^{-1/\nu } \) gives \( \nu \simeq 1.0 \)
and \( h_{c}\equiv h_{c}(\infty )\simeq 0.716 \) (see inset of Fig.
3). This is close to an earlier estimate \( h_{c}\simeq 0.71695 \)
\cite{Ves}, made in a somewhat different version of the model. The
avalanche size distribution has got power laws similar to the BTW
model, at this self-organised critical state at \( h_{av}=h_{c} \).
However the exponents seem to be different \cite{Man91,DD99}, compared
to those of BTW model, for this stochastic model.

\noindent \textbf{b) Simulation studies with pulsed perturbation}

We have considered Manna model on square lattice of different sizes
(\( L= \) \( 100 \), \( 200 \) and \( 300 \)). At a fixed value
of \( L \), for any pile configuration at an average height \( h_{av} \),
a fixed number of heights \( h_{p}=2 \) has been added at any central
point of the stable pile (for which dynamics had stopped). Just after
the addition, the local dynamics starts and it takes a finite time
(iteration number) to return back to the stable state (\( h_{i,j}<2 \)
for all \( (i,j) \)) after several toppling events. For each value
of \( h_{av}(<h_{c}) \) this response time for each pile configuration
has been noted and the average relaxation time \( \tau  \) is obtained
from the average over \( 10^{5} \) different configurations. Near
critical point \( \tau  \) is seen to diverge (see Fig. 4 (a)) as
\( h_{av} \) approaches the critical height \( h_{c} \) with a power
law \( \tau \sim (h_{c}-h_{av})^{-\gamma } \), where \( \gamma \cong 1.2 \).
The plot of \( \tau ^{-1/\gamma } \) with \( h_{av} \) is a straight
line (see inset of Fig. 4(a)) with negative slope. Extrapolating the
straight line and locating the vanishing point of \( \tau ^{-1/\gamma } \)
we have estimated the critical height as \( h_{c}=0.72\pm .01 \),
which is very close to the previous numerical estimate \( h_{c}\cong 0.716 \)
for this model (see inset of Fig. 3).

The size of the damage, i.e., the total number of topplings (after
the addition of pulse) has also been measured for the above cases.
The average (over about \( 10^{5} \) configurations) number of topplings
\( \Delta  \) also diverges as average height \( h_{av} \) approaches
the critical height \( h_{c} \) and near critical point \( \Delta  \)
grows as \( \Delta  \) \( \sim (h_{c}-h_{av})^{-\delta } \), where
\( \delta \cong 2.0 \) (see Fig. 4 (b)). The plot of \( \Delta ^{-1/\delta } \)
versus \( h_{av} \) gives a straight line which can be used to estimate
\( h_{c} \) \( (=0.72\pm .01) \) after extrapolation (see inset
of Fig. 4 (b)).

The correlation length (\( \xi  \)) of the system has been measured
following the same procedure as in the BTW model, described in the
previous section. The average (over about \( 10^{5} \) configurations)
correlation length \( \xi  \) again diverges (see Fig. 4 (c)) as
\( h_{av}\rightarrow  \) \( h_{c} \) and near critical point \( \xi  \)
follows the power law \( \xi  \) \( \sim  \) \( (h_{c}-h_{av})^{-\nu } \),
where \( \nu \cong 1.0 \) . The plot of \( \xi  \)\( ^{-1/\nu } \)
versus \( h_{av} \) is a straight line with negative slope and the
vanishing value of \( \xi  \)\( ^{-1/\nu } \) estimates the critical
density \( h_{c}= \) \( 0.72\pm .01 \) (see inset of Fig. 4 (c))
which is again close to the estimated critical density from direct
numerical study.

\noindent \vskip .2in

\noindent \textbf{IV. Precursors in the random fiber bundle model }

\noindent \textbf{a) The model}

We consider a RFB model containing \( N \) elastic fibers clamped
at two ends, where the failure stress of the individual fibers are
distributed randomly and uniformly within \( 0 \) and \( 1 \) (white
distribution). Global load sharing is assumed and the applied load
on the bundle is democratically shared among the existing intact fibers
of the bundle. With the application of any small load \( F \) (\( = \)
\( \sigma  \) \( N \), with \( \sigma \ll 1 \)) on the bundle,
an initial stress \( \sigma  \) sets in. At the first step, \( \sigma  \)\( N \)
number of fibers are broken off, leaving \( Nu_{1}( \)\( \sigma ) \)
\( = \) \( (1-\sigma ) \)\( N \) number of unbroken fibers. After
this, the applied force is redistributed uniformly among remaining
intact fibers and the stress (per fiber) is then readjusted to a value
\( F/ \)\( [Nu_{1}( \)\( \sigma )] \) \( = \) \( \sigma / \)\( (1-\sigma ) \).
With this new readjusted stress, some extra fibers for which the strengths
are below the above readjusted stress fail and the total number of
broken fibers increases to a value \( N[\sigma / \)\( (1-\sigma )] \),
leaving \( Nu_{2}( \)\( \sigma ) \) \( = \) \( [1-\sigma /(1-\sigma )]N \)
unbroken fibers. This in turn readjusts the stress again and induces
further failure giving rise to a recursive relation: \begin{equation}
\label{qq}
u_{n}(\sigma )=1-\frac{\sigma }{u_{n-1}(\sigma )},
\end{equation}
 for the fraction \( u \) of unbroken fibers at the \( n \)-th and
\( (n-1) \)-th iteration for stress \( \sigma  \). This dynamics
of successive failure propagates therefore in (discrete) time until
\( Nu_{n-1}( \)\( \sigma ) \) \( - \) \( Nu_{n}( \)\( \sigma ) \)
\( \leq  \) \( 1 \), or the successive stress readjustments make
so little change that even one fiber cannot be found in the network
having strength between the successive readjusted value. For an infinite
(\( N\rightarrow \infty  \)) fiber bundle, we denote the fraction
of unbroken fibers here by the fixed point value \( u^{\star }( \)\( \sigma ) \).
The critical stress \( \sigma _{c} \) is determined by that \( \sigma  \)
above which there is no fixed point and \( u_{n}( \)\( \sigma )\rightarrow 0 \)
as \( n\rightarrow \infty  \). Because of the above simple recursion
relation (4) for \( u \), in the uniformly distributed RFB model,
we can easily analyse the asymptotic features of its dynamics. The
differential form of the above recursion relation (4) can be written
as

\noindent \begin{equation}
\label{rr}
\frac{du}{dn}=-\frac{(u^{2}-u+\sigma )}{u}.
\end{equation}

\noindent The fixed point value of \( u \) is obtained by setting
\( du/dn=0 \). This gives

\noindent \begin{equation}
\label{qq}
u^{\star }=\frac{1}{2}+(\sigma _{c}-\sigma )^{1/2},
\end{equation}

\noindent where \( \sigma _{c}=1/4 \). The other root is neglected
here as it is unstable (see equation 7). Expanding the equation (5)
near the fixed point value (6) of \( u \), we can write \( u=u^{\star }+\epsilon  \),
and \begin{equation}
\label{ss}
\frac{d\epsilon }{dn}=-\frac{\epsilon (2u^{\star }-1)}{u^{\star }}\simeq -\epsilon [4(\sigma _{c}-\sigma )^{1/2}],
\end{equation}

\noindent as \( \sigma \rightarrow \sigma _{c} \), which gives \begin{equation}
\label{ww}
u_{n}=u^{\star }+{\textrm{const}}.\exp (-n/\tau _{0}),
\end{equation}

\noindent where

\noindent \begin{equation}
\label{zz}
\tau _{0}=\frac{1}{4}(\sigma _{c}-\sigma )^{-1/2}.
\end{equation}

\noindent \textbf{b) Study of the precursors}

We have simulated the RFB model with a very slow but steady increase
of initial stress \( \sigma  \) on a bundle containing \( N \) fibers
(\( N\sim 10^{8} \)). Application of some small initial stress \( \sigma  \)
(\( = \) \( F/N \)) triggers the dynamics by breaking off a fraction
\( (1-u_{n}) \) of fibers, and global readjustment of the stress
causes further failures (\( u_{n+1}<u_{n} \)). As mentioned before,
after a few steps or iterations, when \( N[u_{n-1}( \)\( \sigma ) \)
\( - \) \( u_{n}( \)\( \sigma )] \) \( \leq  \) \( 1 \), the
dynamics stops and the bundle becomes stable. We note this relaxation
time \( \tau  \) required for the stabilisation. For each (initial)
stress \( \sigma  \) we start afresh with the intact bundle and note
the relaxation time for each \( \sigma  \). The observation continues
until we reach the threshold stress \( \sigma _{c} \) (\( = \) \( 1/4 \)),
above which the bundle fails totally (see Fig. 5). The relaxation
time \( \tau  \) is seen to diverge as \( \sigma  \) \( \rightarrow  \)
\( \sigma _{c} \) following a power law \( \tau  \) \( \sim  \)
\( \tau _{0}\sim  \) \( (\sigma _{c} \) \( - \)\( \sigma )^{-1/2} \)
(see inset of Fig. 5) which can be explained easily using equation
(8).

Similar studies have been made for the breakdown susceptibility \( \chi  \)\( \equiv  \)
\( dm \) \( /d\sigma  \), where \( m \) \( = \) \( N(1- \)\( u^{\star }( \)\( \sigma )) \)
is the total number of fibers broken finally by stress \( \sigma  \)
(see inset of Fig. 5). One finds \( \chi  \) \( \sim  \) \( (\sigma _{c} \)
\( - \)\( \sigma )^{-1/2} \), in agreement with the previous observations
\cite{RS99,PS97}. This can be easily explained from solution (6).

\vskip .2in

\noindent \textbf{V. Summary and concluding remarks}

In all the three dynamical models of failure we have considered here,
we find that long before the occurrence of global failures, the growing
correlations in the dynamics of constituent elements manifest themselves
as various precursors. The number of topplings \( \Delta  \), relaxation
time \( \tau  \) and the correlation length \( \xi  \), in both
BTW and Manna model, grow and diverge following power laws as the
systems approach their respective critical points \( h_{c} \) from
below: \( \Delta  \) \( \sim (h_{c}-h_{av}) \)\( ^{-\delta } \),
\( \tau \sim (h_{c}-h_{av})^{-\gamma } \) and \( \xi  \) \( \sim  \)
\( (h_{c}-h_{av})^{-\nu } \). For two dimensional systems, we find
numerically here \( \delta \simeq 2.0 \), \( \gamma \simeq 1.2 \)
and \( \nu \simeq 1.0 \) for both BTW and Manna model. We could not
thus detect any significant difference in the power laws for these
precursors. We also could not detect any significant finite size effect
in these precursors. Though this size independence of the quantities
we studied look quite unnatural at first sight, there are strong reasons.
Basically, we study the behaviour for \( h_{av}<h_{c} \), the precursor
behaviour, where \( \xi  \) is necessarily finite. As we add here
the tiny pulse at some central site of a relatively large system,
the boundary effect can not be really felt because of the smallness
of \( \xi  \) compared to \( L \) for most values of \( h_{av} \).
This explaines the lack of finite size effect in our precursor studies
(which of course is clearly manifest when we check our model results
at \( h_{av}=h_{c}(L) \)). It may also be noted that since for \( h_{av} \)
near \( h_{c} \), in our system, \( \xi  \) becomes of the order
of \( L \), at \( h_{av}=h_{c}(L) \), our results suggest \( \Delta \sim L^{2.0} \)
and \( \tau \sim L^{1.2} \). This in fact supports the earlier analytic
result for \( h_{av}=h_{c}(L) \) for large but finite systems, as
obtained by Dhar \cite{DD92 99}. Generally, if we write \( \Delta \sim \xi ^{d_{f}} \),
we then get \( d_{f}=\delta /\nu \simeq 2.0 \) for the fractal dimension
of the avalanche clusters.

Apart from the previous attempts \cite{AC96}, an indirect study in
the fixed energy sand pile (FES) model \cite{Ves} also indicated
similar power law behaviour away from the critical point (essentially
for \( h_{av} \) above \( h_{c} \)). In the BTW-FES model, the observed
exponent values for \( \tau  \) and \( \xi  \) differ significantly
from those of ours'. However for the Manna-FES model, these exponent
values are close to our estimates. The FES version of the models are
somewhat different by construction and the discrepancies in case of
BTW-FES estimates (compared to ours') seem to be physical in their
origin. Due to lack of stochasticity, BTW model can stabilise in several
`metastable' states (above \( h_{\textrm{c}} \)) and non-universality
occurs because of different initial conditions. This can also be seen
from the difference in the estimate of the critical point \( h_{c} \)
in the BTW and BTW-FES models; as mentioned before, no such difference
in the \( h_{c} \) estimate seems to exist for the Manna and Manna-FES
models. This difference in the \( h_{c} \) values for the BTW case
might explain the difference in the exponent values we obtained (for
\( h_{av}<h_{c} \)) and those obtained for the corresponding FES
model (for \( h_{av}>h_{c} \)). 

For the random fiber bundle model, we find that the breakdown susceptibility
\( \chi  \) (giving the increment in the number of broken fibers
for an infinitesimal increment of load on the network) and the corresponding
relaxation time \( \tau  \) (required for the network to stabilise,
after successive failures of the fibers), both diverge as the external
load or stress approaches its global failure point \( \sigma _{c} \)
from below: \( \chi  \) \( \sim  \) \( (\sigma _{c} \) \( - \)\( \sigma _{av})^{-1/2} \)
and \( \tau  \) \( \sim  \) \( (\sigma _{c} \) \( - \)\( \sigma _{av})^{-1/2} \).
These results for the RFB model are of course analytically derived
here for uniform distribution of strength of the fibers. It may be
mentioned here that a similar behaviour for the time-to-fracture (for
\( \sigma  \) above \( \sigma _{c} \); diverging with the same exponent
1/2 for \( \tau  \)) was observed in a RFB model where the fibers
relax, under stress, to the elastic strain through viscous damping
\cite{Kun}. However, the relaxational dynamics in this visco-elastic
RFB model is not due to the (self-organising) stress redistributions
among the surviving fibers and, as such, is quite different in its
origin. In fact, this time-to-failure vanishes in the limit of zero
damping coefficient \cite{Kun}. However, the similarities in the
behaviour in such distinctly different situations also indicate interesting
possibilities.

Knowledge of the precursors and their power laws should help estimating
precisely the location of the global failure or critical point from
the proper extrapolation of the above quantities, which are available
long before the failure occurs. The usefulness of such precursors
can hardly be overemphasized.

\vskip .1in

\noindent \textbf{Acknowledgement:} We are grateful to P. Bhattacharyya,
A. Chakraborti, R. Karmakar, F. Kun, S. S. Manna and A. Vespignani
for useful comments, suggestions and discussions.

\noindent \textbf{Figure captions}

\noindent \vskip .1in

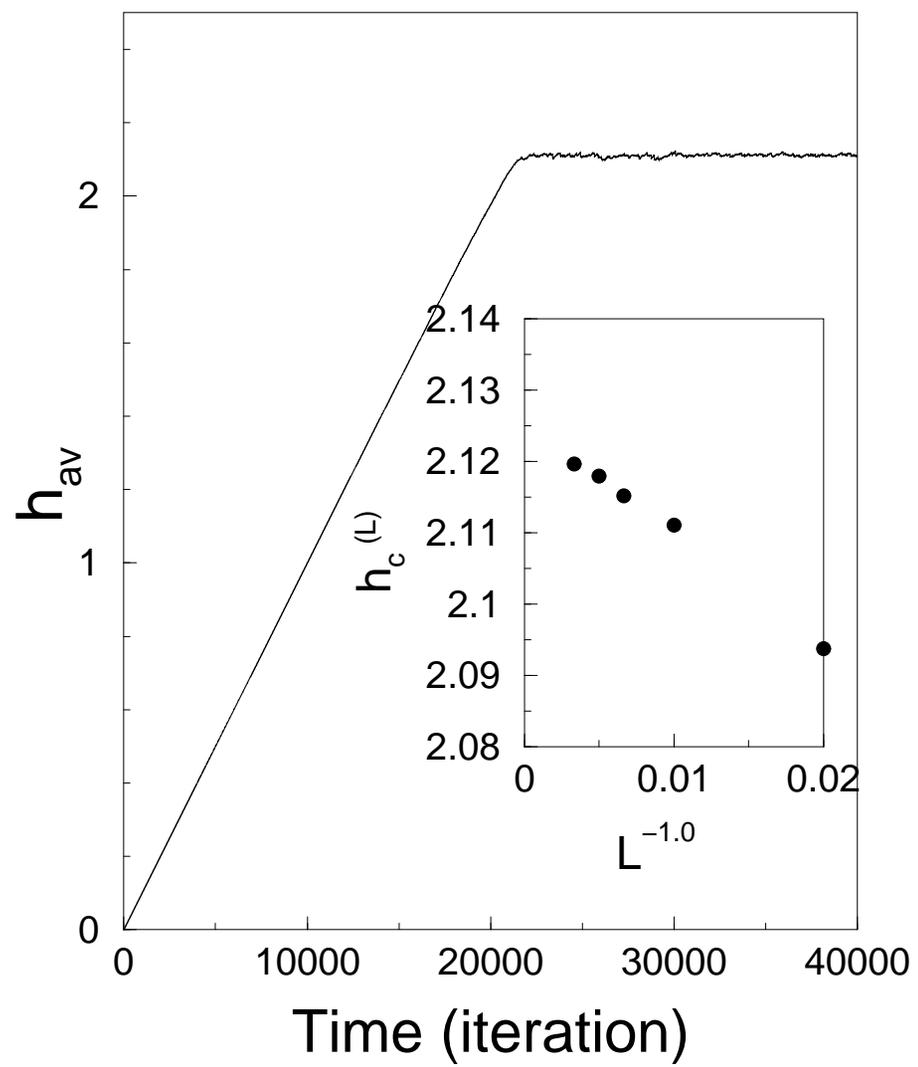
\noindent \textbf{Fig. 1:} The growth of average height \( h_{av} \)
(\( <h_{c}(L) \)) of the BTW model against the number of iterations
of adding unit heights \( (L=100) \) . In the inset, we show the
finite size behaviour of the critical height \( h_{c}(L) \), obtained
from simulation results for different \( L \).

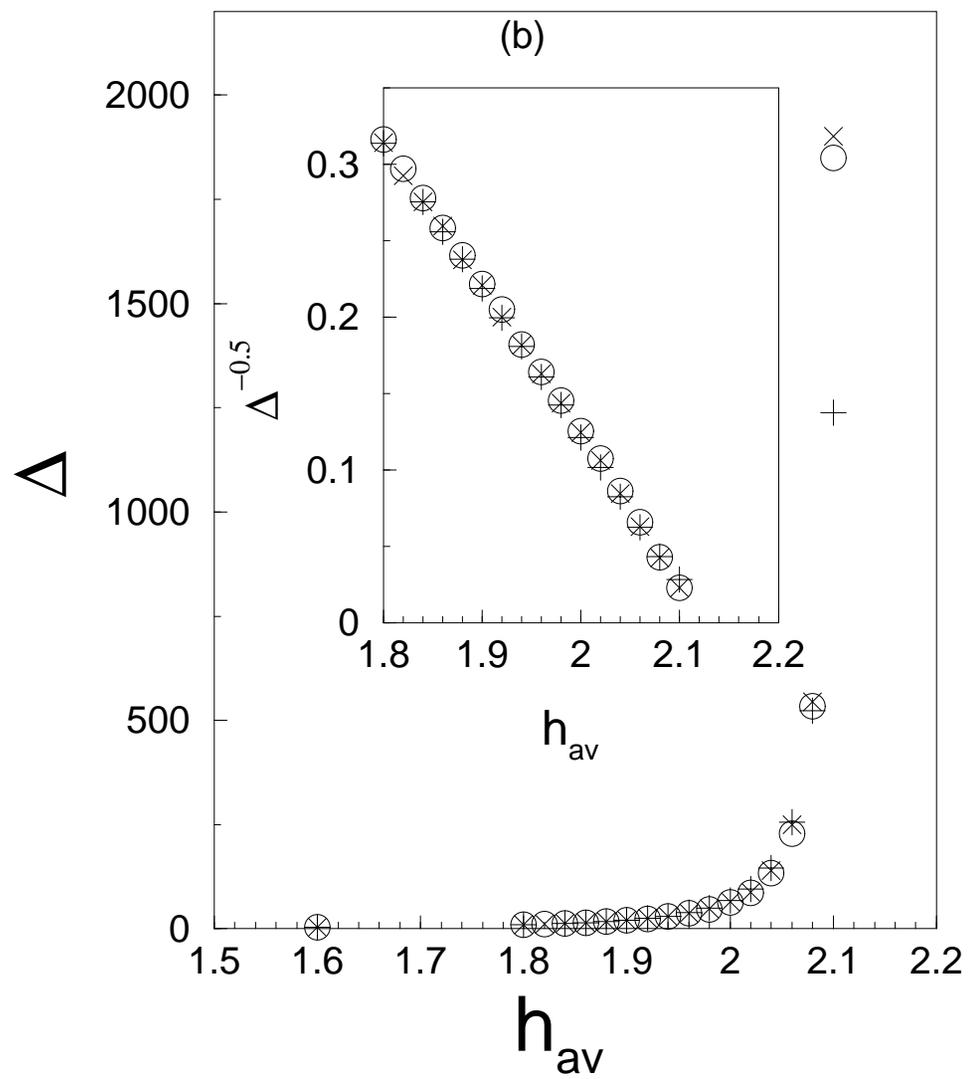
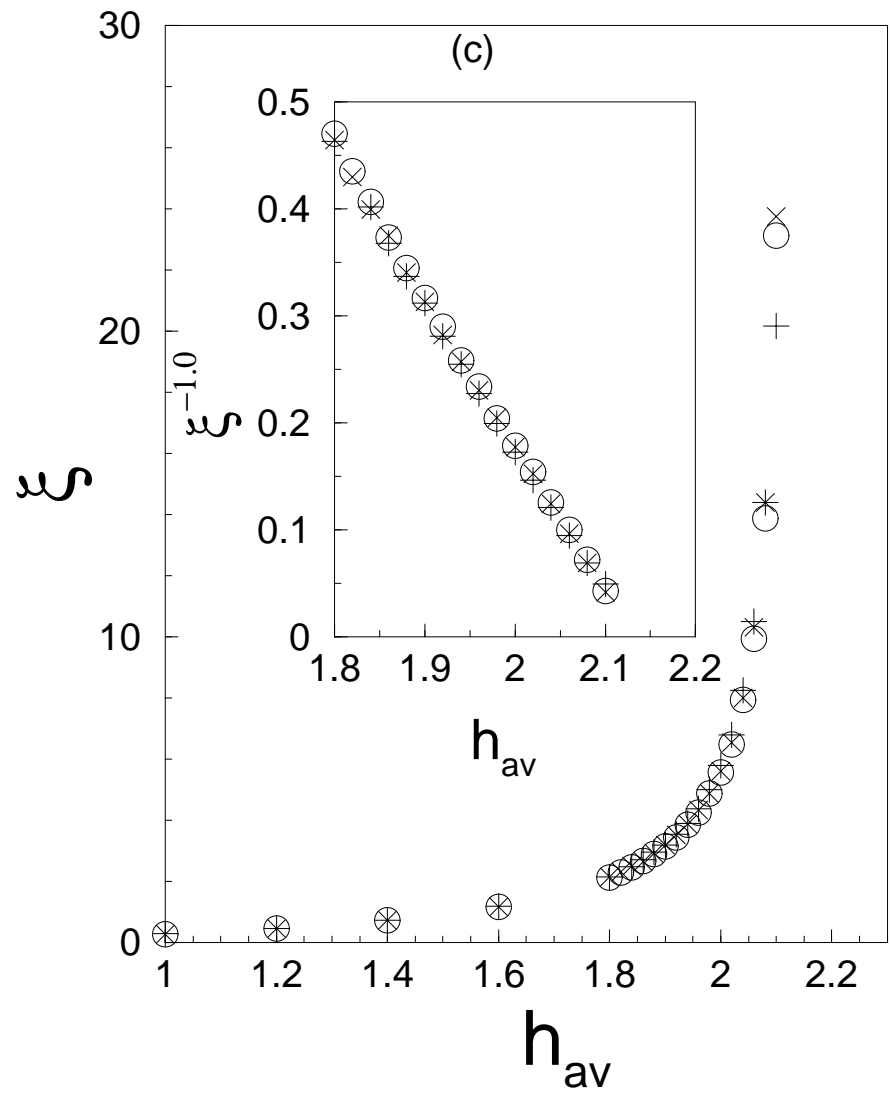
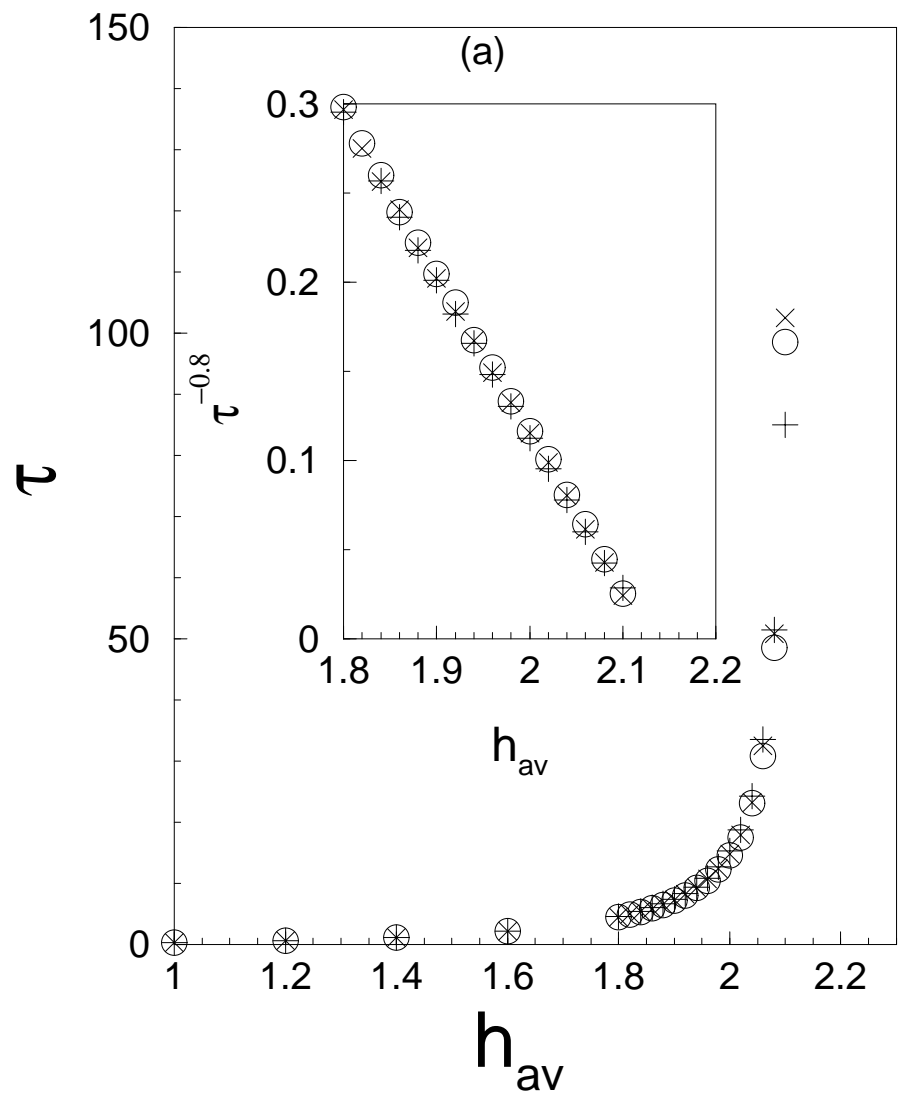
\noindent \textbf{Fig. 2:} The variations of the precursors with \( h_{av} \)
(\( <h_{c}(L) \)) in the BTW model for different system sizes: \( L=100 \)
(plus) \( L=200 \) (cross) and \( L=300 \) (open circle). (a) For
relaxation time \( \tau  \); in the inset \( \tau ^{-0.8} \) is
plotted against \( h_{av} \). (b) For the total number of topplings
\( \Delta  \); inset shows \( \Delta ^{-0.5} \) versus \( h_{av} \)
plot. (c) For the correlation length \( \xi  \); in the inset, \( \xi ^{-1.0} \)
is plotted against \( h_{av} \).

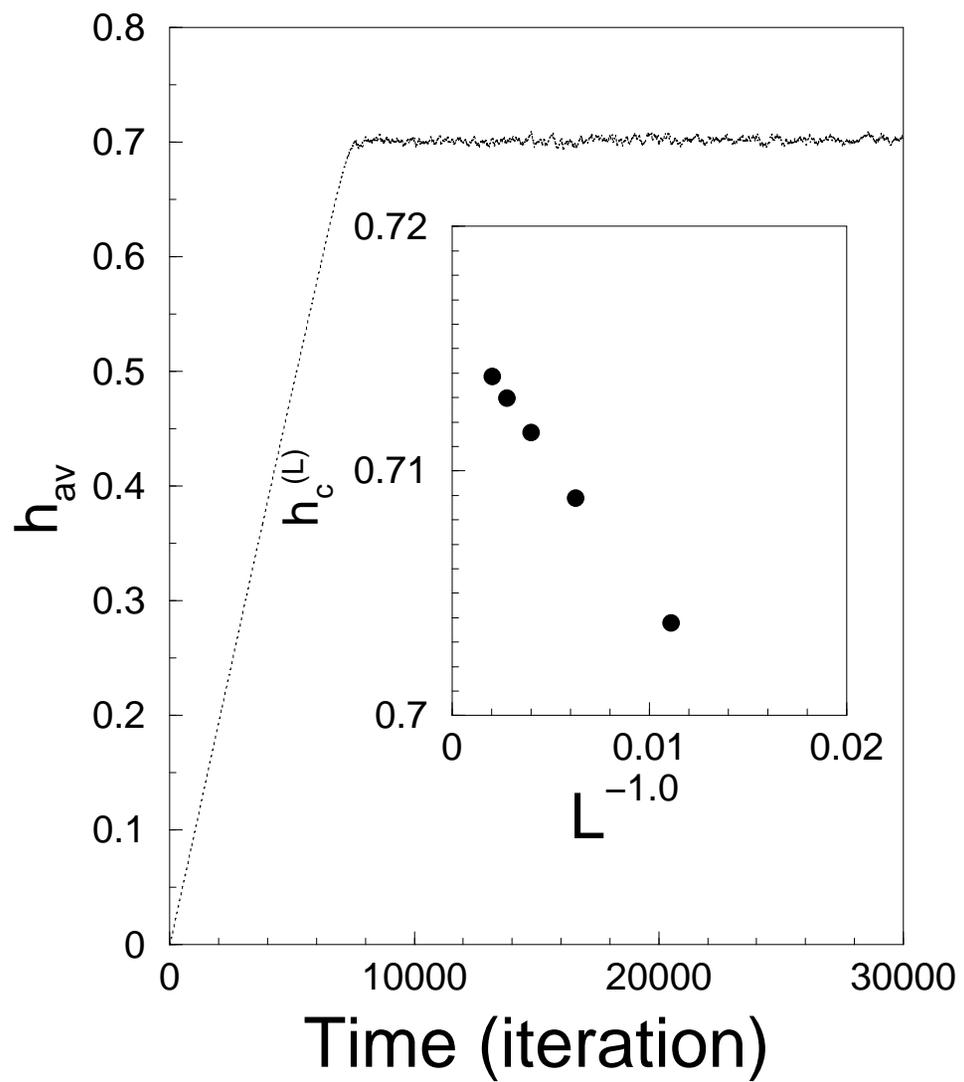
\noindent \textbf{Fig. 3}: The growth of average height \( h_{av} \)
(\( <h_{c}(L) \)) of the Manna model against the number of iterations
of adding unit heights \( (L=100) \) . In the inset, we show the
finite size dependence of the critical height \( h_{c}(L) \), obtained
from simulation results for different \( L \).

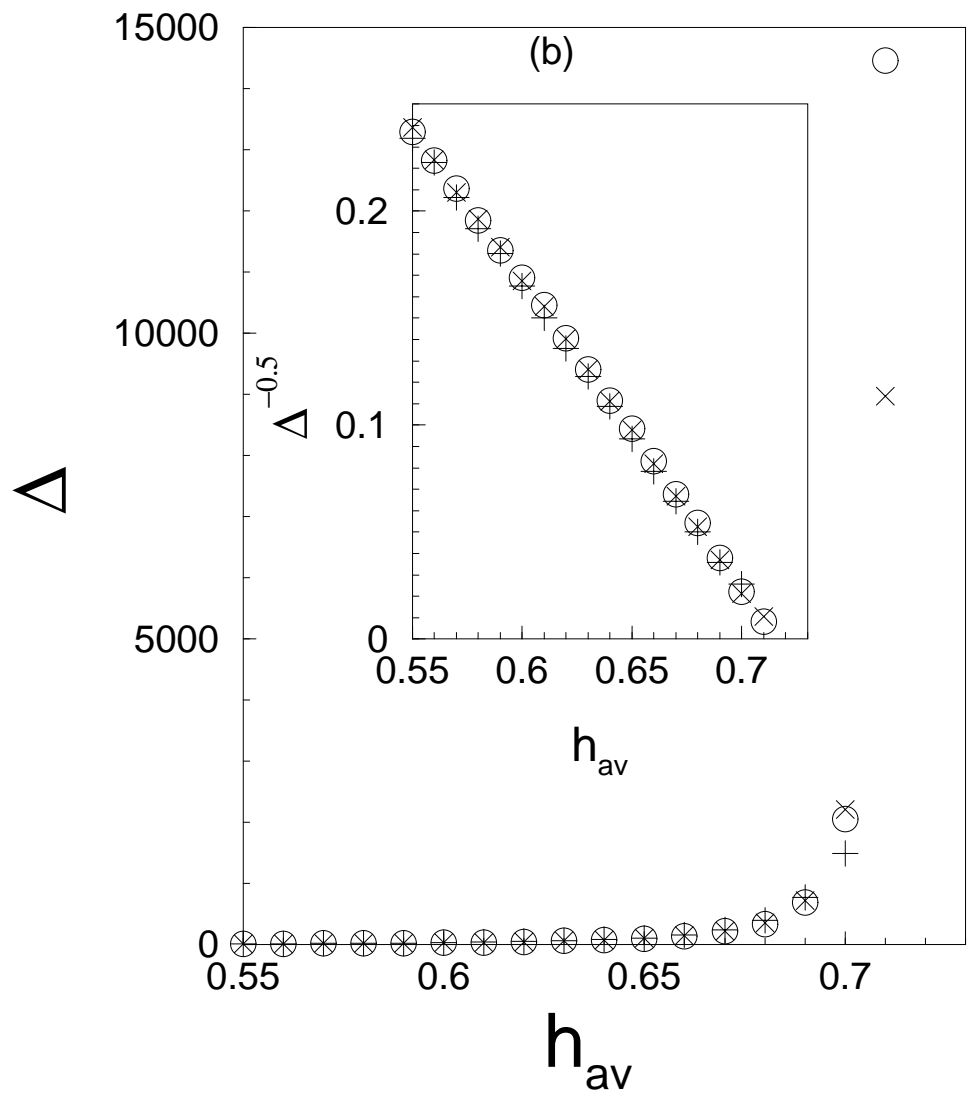
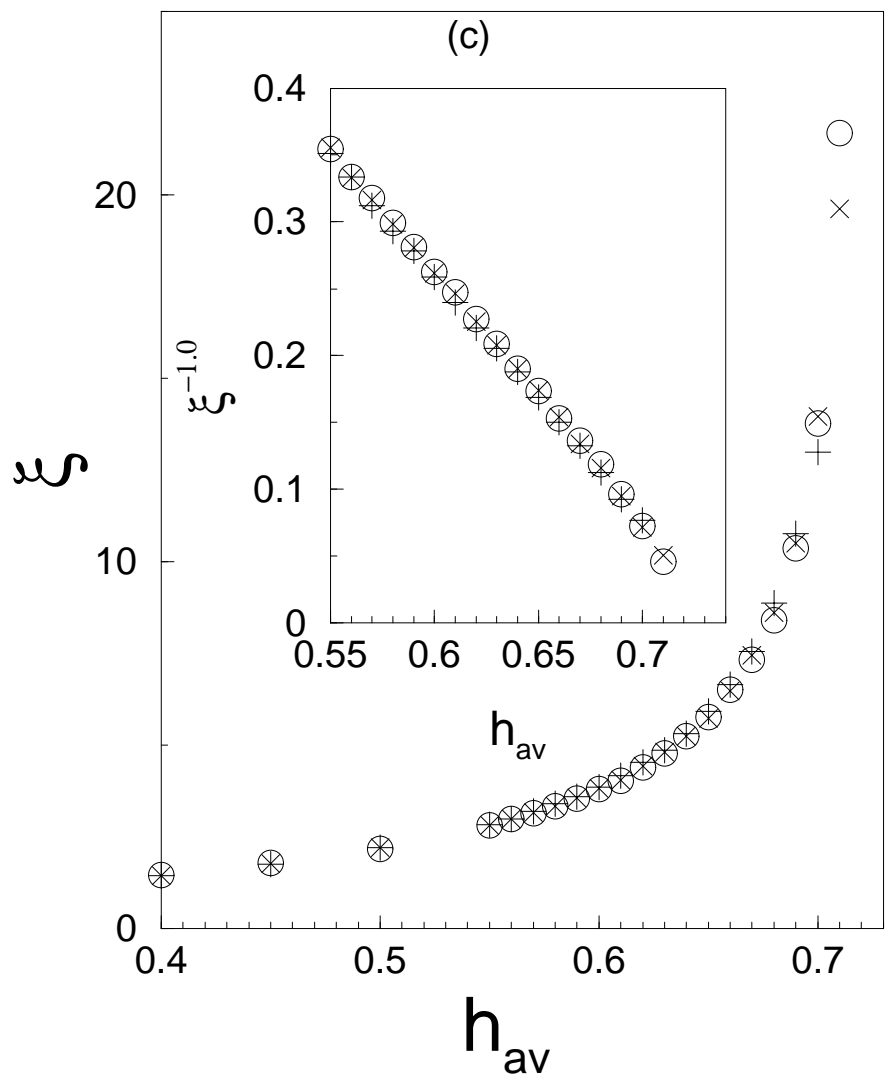
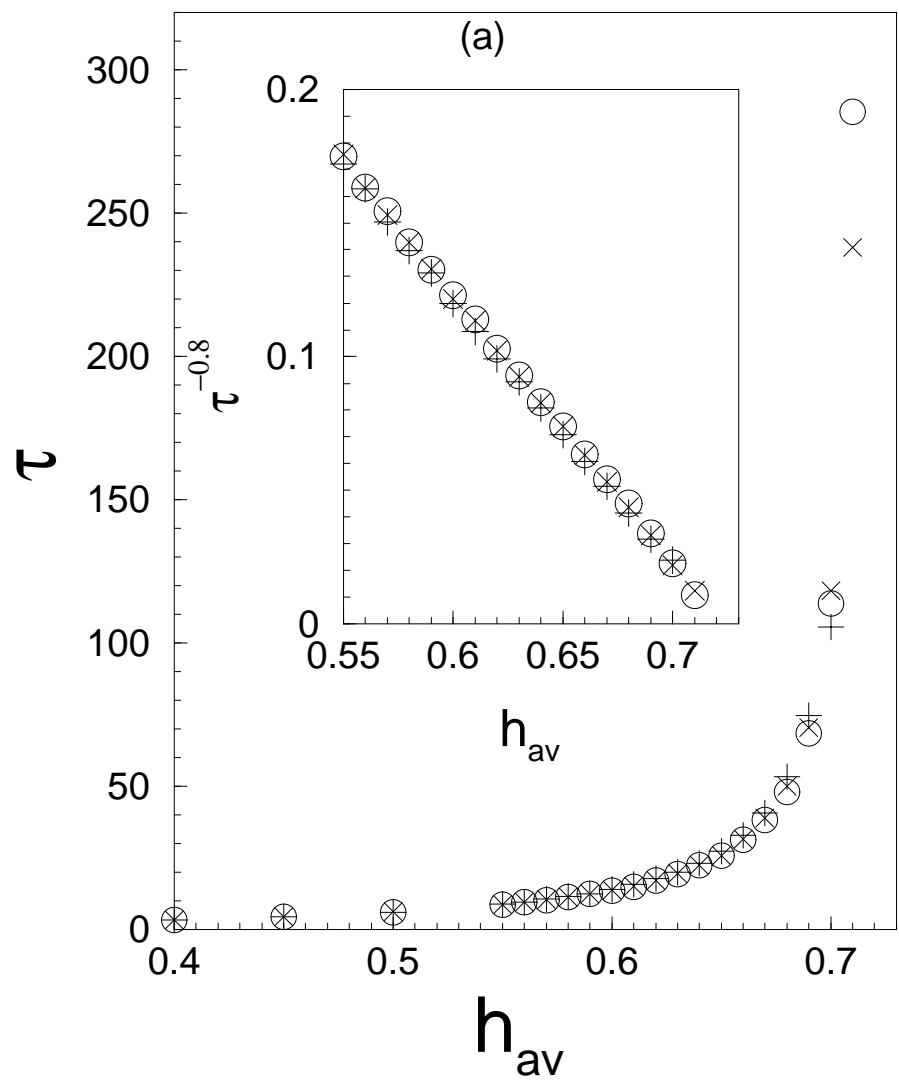
\noindent \textbf{Fig. 4:} The variations of the precursors with \( h_{av} \)
(\( <h_{c}(L) \)) in the Manna model for different system sizes:
\( L=100 \) (plus) \( L=200 \) (cross) and \( L=300 \) (open circle).
(a) For relaxation time \( \tau  \); in the inset \( \tau ^{-0.8} \)
is plotted against \( h_{av} \). (b) For the total number of topplings
\( \Delta  \); inset shows \( \Delta ^{-0.5} \) versus \( h_{av} \)
plot. (c) For the correlation length \( \xi  \); in the inset, \( \xi ^{-1.0} \)
is plotted against \( h_{av} \).

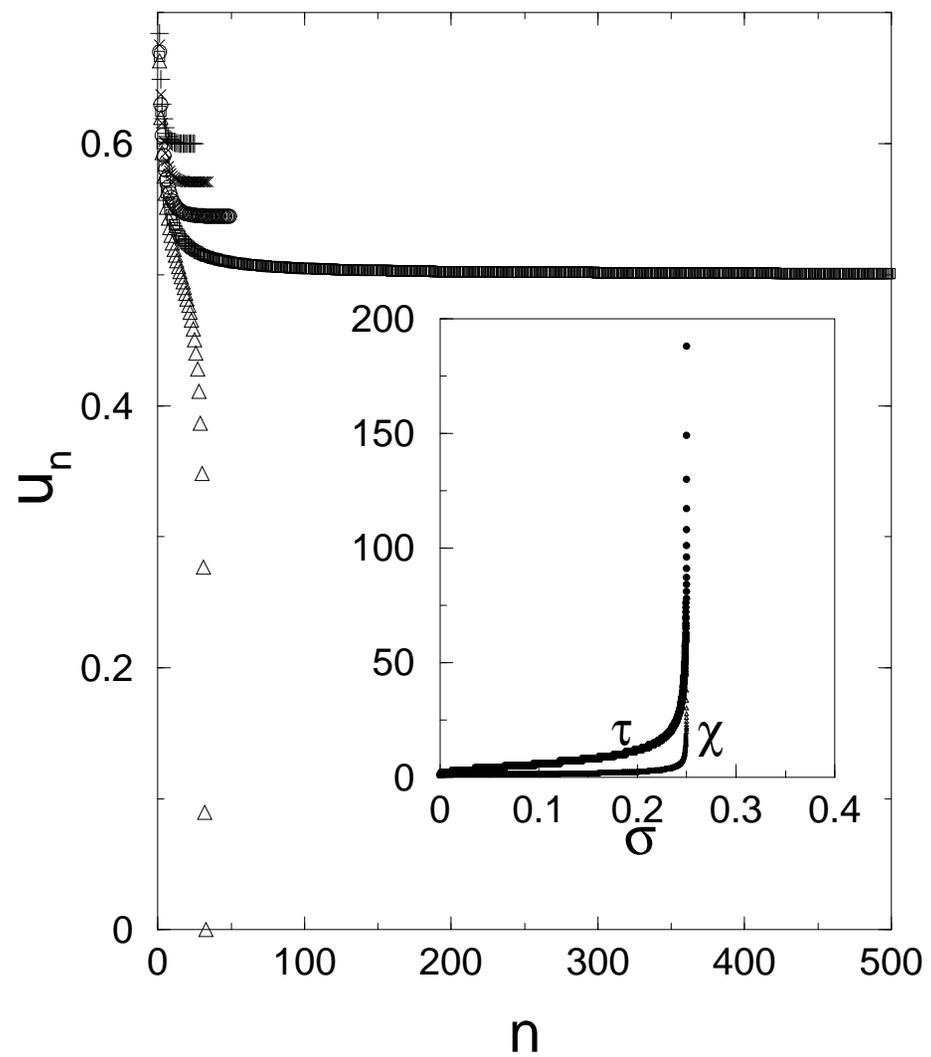
\noindent \textbf{Fig. 5}: Fraction of the unbroken fibers \( u_{n} \)
at different times or iterations \( n \) in a RFB model with uniform
strength distribution, for different values of (initial) stress: \( \sigma =0.24 \)
(plus), \( \sigma =0.245 \) (cross), \( \sigma =0.248 \) (open circle),
\( \sigma =0.25 \) (open square), \( \sigma =0.252 \) (open triangle).
Note that the last value of \( \sigma  \) is greater than \( \sigma _{c}(=1/4) \),
and the fraction of unbroken fibers goes to zero here. Inset shows
how the susceptibility \( \chi  \) (up triangle) and the relaxation
time \( \tau  \) (filled circle) both diverge as \( \sigma \rightarrow \sigma _{c}(=1/4) \). 

\noindent \vskip .1in

\begin{thebibliography}{10}
\bibitem[1]{BTW187}P. Bak, C. Tang and K. Wiesenfeld, Phys. Rev. Lett. \textbf{59} 381
(1987). 
\bibitem[2]{BTW287}P. Bak, C. Tang and K. Wiesenfeld, Phys. Rev. A. \textbf{38} 364 (1988). 
\bibitem[3]{GM90}P. Grassberger and S. S. Manna, J. Phys. France \textbf{51} 1077 (1990). 
\bibitem[4]{B2000}P. Bak, \emph{How Nature Works,} Oxford Univ. Press, Oxford (1997). 
\bibitem[5]{Man91}S. S. Manna, J. Phys. A: Math. Gen. \textbf{24} L363 (1991). 
\bibitem[6]{DD99}D. Dhar, Physica A \textbf{270} 69 (1999). 
\bibitem[7]{Dan45}H. E. Daniels, Proc. R. Soc. London A \textbf{183} 405 (1945). 
\bibitem[8]{HH92}P. C. Hemmer and A. Hansen, J. Appl. Mech. \textbf{59} 909 (1992). 
\bibitem[9]{HH94}A. Hansen and P. C. Hemmer, Phys. Lett. A \textbf{184} 394 (1994). 
\bibitem[10]{D92}D. Sornette, J. Phys. I (France) \textbf{2} 2089 (1992). 
\bibitem[11]{RS99}R. da Silveria, Am. J. Phys. \textbf{67} 1177 (1999). 
\bibitem[12]{Pach2000}Y. Moreno, J. B. Gomez and A. F. Pacheco, Phys. Rev. Lett. \textbf{85}
2865 (2000). 
\bibitem[13]{BB97}B. K. Chakrabarti and L. G. Benguigui, \textit{Statistical Physics
of Fracture and Breakdown in Disorder Systems}, Oxford Univ. Press,
Oxford (1997). 
\bibitem[14]{DT99}D. L. Turcotte, Rep. Prog. Phys. \textbf{62} 1377 (1999). 
\bibitem[15]{AC96}M. Acharyya and B. K. Chakrabarti, Physica A \textbf{224} 254 (1996);
Phys. Rev. E \textbf{53} 140 (1996). 
\bibitem[16]{PS97}S. Zapperi, P. Ray, H. E. Stanley and A. Vespignani, Phys. Rev. Lett.
\textbf{78} 1408 (1997). 
\bibitem[17]{Manna90}S. S. Manna, J. Stat. Phys. \textbf{59} 509 (1990). 
\bibitem[18]{Ves}A. Vespignani, R. Dickman, M. A. Munoz and S. Zapperi, Phys. Rev.
E \textbf{62} 4564 (2000); A. Chessa, E. Marinari and A. Vespignani,
Phys. Rev. Lett. \textbf{80} 4217 (1998). 
\bibitem[19]{DD92 99}D. Dhar, Physica A \textbf{186} 82 (1992); D. Dhar, Physica A \textbf{263}
4 (1999). 
\bibitem[20]{Kun}R. C. Hidalgo, F. Kun and H. J. Herrmann, cond-mat/0103232. 
\end{thebibliography}
\end{document}